\documentclass[a4paper]{jpconf}
\usepackage{graphicx}
\usepackage{iopams}
\begin{document}
\title{The Dicke model as the contraction limit of a pseudo-deformed Richardson-Gaudin model}

\author{Pieter~W.\ Claeys\textsuperscript{1,2}, Stijn~De~Baerdemacker\textsuperscript{1,2}, Mario~Van~Raemdonck\textsuperscript{1,2,3}, Dimitri~Van~Neck\textsuperscript{1,2}}

\address{\textsuperscript{1} Ghent University, Center for Molecular Modeling, Technologiepark 903, 9052 Ghent, Belgium\\
\textsuperscript{2} Ghent University, Department of Physics and Astronomy, Proeftuinstraat 86, 9000 Ghent, Belgium\\
\textsuperscript{3}Ghent University, Department of Inorganic and Physical Chemistry, Krijgslaan 281 (S3), 9000 Ghent, Belgium }

\ead{stijn.debaerdemacker@ugent.be}

\begin{abstract}
The Dicke model is derived in the contraction limit of a pseudo-deformation of the quasispin algebra in the $su(2)$-based Richardson-Gaudin models.  Likewise, the integrability of the Dicke model is established by constructing the full set of conserved charges, the form of the Bethe Ansatz state, and the associated Richardson-Gaudin equations.  Thanks to the formulation in terms of the pseudo-deformation, the connection from the $su(2)$-based Richardson-Gaudin model towards the Dicke model can be performed adiabatically.
\end{abstract}

\section{Introduction}

The interaction of a single quantized mode of electromagnetic radiation (photons) with a two-state system, such as a nuclear spin or two-level atom, can be modeled by means of the Rabi Hamiltonian \cite{grynberg:2010}.  Although simple and physically transparent in its formulation, the Rabi Hamiltonian does not offer a known exact eigenstate, due to the `counter-rotating' interaction terms in the Hamiltonian.  In the resonance regime however, these fast-frequency counter-rotating terms average out with respect to the collective two-level oscillations, and can be neglected in first-order perturbation (the so-called Rotating Wave Approximation (RWA)).  This gives rise to the Jaynes-Cummings (JC) Hamiltonian of quantum electrodynamics \cite{jaynes:1963}.  Due to the RWA, the JC Hamiltonian exhibits an additional symmetry, which allows for the decomposition of the Hilbert space into irreducible representations (irreps) conserving the total number of (bosonic and atomic) excitations.  For the two-level ($s=1/2$ quasispin) formulation of the JC model, these irreps remain two-dimensional and therefore reduce to two-level mixing models.  One of the successes of theoretical and experimental quantum mechanics is that such simple theoretical models have been validated in sophisticated experiments situated in cavity quantum electrodynamics \cite{brune:1996}, circuit quantum electrodynamics \cite{wallraff:2004}, etc.

A direct consequence of the additional symmetry is that the two-level JC model exhibits two linearly independent conserved operators, \emph{i.e.}\ the Hamiltonian and the operator counting the total number of excitations.  As the model has two degrees of freedom (the electromagnetic photon mode and the two-level system), the model is \emph{integrable} by definition \cite{caux:2011}, and supports a Bethe Ansatz product wavefunction \cite{gaudin:1976}.  This result can be generalized to general $(2s+1)$-level systems, also referred to as Tavis-Cummings models \cite{tavis:1968}, or a collections of inequivalent $(2s+1)$-level systems interacting via a single mediating bosonic mode, called the Dicke model \cite{dicke:1954}.  The integrability of the Dicke model has been established first by Gaudin \cite{gaudin:1976}, by means of an infinite dimensional Schwinger representation of the central spin in the Gaudin magnet.  Later, the full set of conserved charges of the Dicke model has been derived by Dukelsky et.\ al. \cite{dukelsky:2004b}, by mapping one of the $su(2)$ quasispin copies of the Richardson-Gaudin conserved charges onto the bosonic mode \cite{ortiz:2005}.  As a Bethe Ansatz integrable model, the conserved charges and Bethe Ansatz state of the Dicke model can also be obtained from the Algebraic Bethe Ansatz method \cite{babelon:2007}, however a direct and simplified derivation of the Bethe-Ansatz state with the corresponding Bethe equations solution is feasible using a commutator scheme \cite{tsyplyatyev:2010}.

For the derivation of the conserved charges of the Dicke model from those of the Richardson-Gaudin model \cite{gaudin:1976,dukelsky:2004b}, a contraction of one of the quasispin algebras in the model was required.  This contraction maps the $su(2)$ quasispin directly onto a bosonic Heisenberg-Weyl algebra $hw(1)$.   Recently, it has been shown how the Bethe Ansatz states in the Richardson-Gaudin (RG) solution \cite{richardson:1963,richardson:1964a} of the reduced Bardeen-Cooper-Schriefer (BCS) Hamiltonian for conventional superconductivity \cite{bardeen:1957} can be connected adiabatically to a product state of generalized bosons \cite{debaerdemacker:2012b} employing a pseudo deformation of the quasispin algebra.  The product structure of the Bethe Ansatz state is reminiscent of the projected BCS approximation, in which the superconducting state is approximated by a condensate of collective Cooper pairs \cite{bardeen:1957,cooper:1956}.  However, in general the RG variables (or rapidities) parametrizing the generalized quasispin creation operators in the Bethe Ansatz state differ from each other, which is not reconcilable with the concept of a true condensate.  The pseudo deformation maps the hard-core bosonic $su(2)$ quasispin algebra of the BCS pairing Hamiltonian into a genuinely bosonic $hw(1)$ by means of a continuous pseudo-deformation parameter.  The reduced BCS Hamiltonian remains integrable along the path of deformation, allowing for an adiabatic and injective mapping of the exact Bethe Ansatz states into a condensate of (orthogonal) bosonic modes \cite{debaerdemacker:2011}.  Accordingly, the coupled set of RG equations of integrability reduce to a single decoupled equation, equivalent to the secular equation of the particle-particle Tamm-Dancoff Approximation ($pp$-TDA) for the elementary pairing modes for the reduced BCS Hamiltonian \cite{ring:2004}.  Reversely, the pseudo deformation enables one to numerically reconstruct the solution of the coupled set of non-linear RG equations from the simpler decoupled secular $pp$-TDA equation, by adiabatically reintroducing the Pauli principle in the integrable model \cite{debaerdemacker:2012b,vanraemdonck:2014}.

In the present manuscript, we will reassess the derivation of the conserved charges of the Dicke from those of the Richardson-Gaudin (RG) models in the framework of the pseudo-deformed  quasispin algebra.  This will enable an adiabatic connection of the Dicke model with the $su(2)$ based Richardson-Gaudin models.  The connection can be made consistently on the level of the Hamiltonian, the conserved charges, the Bethe Ansatz state, as well as the Bethe (or Richardson-Gaudin) equations, and sheds light on how the Dicke model can be embedded within the larger class of Richardson-Gaudin integrable models.

\section{Richardson-Gaudin models}

The conserved charges of the Richardson-Gaudin \cite{ortiz:2005} models are parametrized by
\begin{equation}\label{rg:conservedcharges}
R_i=S_i^0 + g\sum_{k\neq i}^m[X_{ik}\case{1}{2}(S_k^\dag S_i + S^\dag_i S_k) + Z_{ik} S_i^0 S_k^0]
\end{equation}
with the set of operators $\{S_i^\dag,S_i,S_i^0\}$ ($i=1\dots m$) spanning a set of $m$ mutually independent $su(2)_i$ quasispin algebras 
\begin{equation}
[S_i^0,S^\dag_k]=\delta_{ik}S^\dag_k,\qquad[S_i^0,S_k]=-\delta_{ik}S_k,\qquad[S_i^\dag,S_k]=2\delta_{ik}S_k^0,
\end{equation}
with irreps $|s_i,\mu_i\rangle$, and the $X$ and $Z$ antisymmetric matrices obeying the Gaudin algebra
\begin{equation}\label{rg:gaudinalgebra}
X_{ij}X_{jk}-X_{ik}(Z_{ij}+Z_{jk})=0,\qquad \forall i\neq j\neq k\neq i.
\end{equation}
As soon as $X$ and $Z$ span the Gaudin algebra, the conserved charges commute mutually
\begin{equation}
[R_i,R_j]=0,\qquad\forall i,j,
\end{equation}
and therefore define a RG integrable model.  Because the conserved charges (\ref{rg:conservedcharges}) are in involution, they have a common set of eigenstates.  These eigenstates are given by means of the Bethe Ansatz
\begin{equation}\label{rg:betheansatz}
|\psi\rangle=\prod_{\alpha=1}^N\left(\sum_{i=1}^mX_{i\alpha}S_i^\dag\right)|\theta\rangle
\end{equation}
with $N$ the number of excitations, and $|\theta\rangle=\otimes_{i=1}^m|s_i,-s_i\rangle$ the tensor product of the lowest-weight irreps of each $su(2)_i$ copy.  The notation $X_{i\alpha}$ signifies that the $m$-dimensional matrix $X$ has been extended to an $(m+N)$-dimensional matrix which still obeys the Gaudin algebra (\ref{rg:gaudinalgebra}) and the antisymmetry condition.  For state (\ref{rg:betheansatz}) to be an eigenstate of the conserved charges (\ref{rg:conservedcharges}), the extended $Z$ (and $X$) matrices need to satisfy the following Bethe ansatz equations 
\begin{equation}\label{rg:rgequations}
1+g\sum_{i=1}^m Z_{i\alpha}s_i-g\sum_{\beta\neq\alpha}Z_{\beta\alpha}=0,\qquad\forall \alpha=1\dots N.
\end{equation}
Up to this point, the Gaudin algebra has not been specified, so these results are fully independent of the particular realisation of $X$ and $Z$.  There are several parametrizations existing in the literature \cite{dukelsky:2001, ortiz:2005}, which can be classified according to the following copy-independent relation
\begin{equation}
X_{ij}^2-Z_{ij}^2=c,\qquad \forall i\neq j,
\end{equation}
with $c$ a constant for all $i\neq j$.  The widely used $c=0$ parametrization gives rise to the rational model, whereas the $c>0$ and $c<0$ are known as the trigonometric and hyperbolic models respectively\footnote{although the trigonometric and hyperbolic case can be transformed into one another by introducing an imaginary coupling constant $g\rightarrow ig$}.  In the present manuscript, we will mainly focus on the trigonometric case, which can be parametrized in terms of a set of $m$ real variables $\{\eta_i\}$  \cite{ortiz:2005}
\begin{equation}\label{rq:trigonometric}
X_{ij}=\frac{\sqrt{(1+\eta_i^2)(1+\eta_j^2)}}{\eta_i-\eta_j},\quad Z_{ij}=\frac{1+\eta_i\eta_j}{\eta_i-\eta_j}
\end{equation}
and $c=1$, but equivalent parametrizations exist \cite{dukelsky:2004b}.
\section{Pseudo deformed quasi-spin algebra}
The pseudo deformed quasispin algebra $su(2)$ is given by \cite{debaerdemacker:2012b}
\begin{equation}
[S^0(\xi),S^\dag(\xi)]=S^\dag(\xi),\quad[S^0(\xi),S(\xi)]=-S(\xi),\quad[S^\dag(\xi),S(\xi)]=2\xi S^0(\xi) +(\xi-1)\case{1}{2}\Omega
\end{equation}
with $\xi$ the pseudo-deformation parameter and $\Omega=2s+1$ the dimension (or degeneracy) of the original quasispin irrep ($\xi=1$).  It can be easily verified that the limits $\xi=1$ and $\xi=0$ give rise to the original $su(2)$ algebra and a (unnormalised) bosonic $hw(1)$ algebra respectively.  The nomenclature \emph{pseudo} deformation is chosen because this algebra can be reduced to a canonical $su(2)$ algebra
\begin{equation}
[A^0(\xi),A^\dag(\xi)]=A^\dag(\xi),\qquad[A^0(\xi),A(\xi)]=-A(\xi),\qquad[A^\dag(\xi),A(\xi)]=2A^0(\xi),
\end{equation}
with 
\begin{equation}
A^\dag(\xi)=\frac{1}{\sqrt{\xi}}S^\dag(\xi),\quad A(\xi)=\frac{1}{\sqrt{\xi}}S(\xi),\quad A^0(\xi)=S^0(\xi)+\left(1-\frac{1}{\xi}\right)\frac{1}{4}\Omega
\end{equation}
except for the $\xi=0$ limit.  For this limit, the following operators
\begin{equation}
b^\dag=\sqrt{\case{2}{\Omega}}S^\dag(0),\quad b^\dag=\sqrt{\case{2}{\Omega}}S^\dag(0),\quad b^\dag b=S^0(0)
\end{equation}
close the $hw(1)$ algebra
\begin{equation}
[b^\dag b,b^\dag]=b^\dag,\quad [b^\dag b,b]=-b\quad [b,b^\dag]=1.
\end{equation}
The representations of the $\{A^\dag(\xi),A(\xi),A^0(\xi)\}$  algebra are labeled by $s(\xi)=s(1)+(\frac{1}{\xi}-1)\Omega$.  The physical interpretation of this is that the effective degeneracy of the quasispin algebra is gradually increased with decreasing $\xi$, reaching infinity for the full $\xi=0$ contraction limit.  It should be noted that only discrete values of $\xi_n=\frac{2\Omega}{n+2\Omega}$ (with $n\in\mathbb{N}$) give rise to unitary irreps.  Nevertheless, this is not an obstacle because the theory of RG integrability does not depend on matrix representations (with integer dimensions), and the parameter $\xi$ can be regarded as a continuous variable.  This is illustrated by the following construction.  Because the set of generators $\{A^\dag(\xi),A(\xi),A^0(\xi)\}$ span an $su(2)$ algebra, the following set of conserved charges are equally in involution
\begin{equation}\label{pseudodeformation:conservedcharges}
R_i(\xi)=A_i^0(\xi) + g\xi\sum_{k\neq i}^m[X_{ik}\case{1}{2}(A_k^\dag(\xi) A_i(\xi) + A^\dag_i(\xi) A_k(\xi)) + Z_{ik} A_i^0(\xi) A_k^0(\xi)],
\end{equation}
provided the matrices $X$ and $Z$ fulfill the Gaudin algebra (\ref{rg:gaudinalgebra}).  Note that the coupling constant $g$ in eq.\ (\ref{rg:conservedcharges}) has been renormalized as $g\xi$.  The Bethe Ansatz state 
\begin{equation}
|\psi\rangle=\prod_{\alpha=1}^N\left(\sum_{i=1}^mX_{i\alpha}A_i^\dag(\xi)\right)|\theta\rangle
\end{equation}
is again an eigenstate of the conserved charges (\ref{pseudodeformation:conservedcharges}) if the pseudo-deformed RG equations
\begin{equation}\label{pseudodeformation:rgequations}
1+g\sum_{i=1}^m Z_{i\alpha}\xi s_i(\xi)-g\xi\sum_{\beta\neq\alpha}Z_{\beta\alpha}=0,\qquad\forall \alpha=1\dots N,
\end{equation}
are solved.  Although the irrep labels $s(\xi)$ approach infinity for $\xi\rightarrow0$, the value of $\xi s(\xi)$ remains finite over the range $\xi\in[0,1]$ with $\lim_{\xi\rightarrow0}[\xi s(\xi)]=\Omega$.  Consequently, the problem of solving the set of pseudo-deformed RG equations (\ref{pseudodeformation:rgequations}) is perfectly well defined for any value of the parameter $\xi\in[0,1]$, and is not restricted to discrete values $\xi_n=\frac{2\Omega}{n+2\Omega}$ for unitary irreps.  The $\xi\rightarrow0$ limit offers a particular case, because the conserved charges (up to a diverging constant) become purely bosonic
\begin{equation}
R_i(\xi=0)=b_i^\dag b_i +g\sum_{k\neq i}^m[\case{1}{4}X_{ik}\sqrt{\Omega_i\Omega_k}(b_i^\dag b_k+b_k^\dag b_i)-\case{1}{4}Z_{ik}(\Omega_ib_k^\dag b_k +\Omega_k b_i^\dag b_i)]
\end{equation}
and the pseudo-deformed RG equations become completely decoupled
\begin{equation}\label{pseudodeformation:rgequations:contractionlimit}
1+g\sum_{i=1}^m Z_{i\alpha}\Omega_i=0,\qquad\forall \alpha=1\dots N,
\end{equation}
which is equivalent to a set of independent secular equations of the $pp$-TDA.  This construction forms the backbone of the numerical RG solver method introduced in \cite{debaerdemacker:2012b,vanraemdonck:2014}.  In this approach, the RG equations are solved in the tractable full contraction limit (\ref{pseudodeformation:rgequations:contractionlimit}), and are then adiabatically brought into the original form (\ref{rg:rgequations}) by tuning $\xi\rightarrow1$.
\section{Dicke model derived from Richardson-Gaudin models}
The pseudo-deformation in the previous section was used to transform all $su(2)_i$ copies ($i\in[1\dots m]$) in the system into a bosonic $hw(1)_i$.  Because the JC and Dicke models contain only one bosonic mode, it would be interesting to see whether an equivalent scheme can be used to obtain the Dicke model from the $\otimes_{i=0}^msu(2)$ RG systems by deforming a single quasispin copy.  Denote this special copy by $i=0$, and consider the conserved charge
\begin{equation}\label{dicke:conservedcharge0}
R_0(\xi)=A_0^0(\xi)+g\sum_{k\neq0}^m[\case{1}{2}X_{0k}(A_0^\dag(\xi)S_k + S_k^\dag A_0(\xi))+Z_{0k}A_0^0(\xi) S_k^0].
\end{equation}
In contrast to the pseudo-deformed conserved charges (\ref{pseudodeformation:conservedcharges}), the coupling constant will be renormalized as $g=\sqrt{\frac{8\xi}{\Omega_0 G^2}}\frac{G^2}{\hbar\omega}$, with $G$ a finite, $\xi$-independent, constant and $\hbar\omega$ the energy of the electromagnetic field.  The other conserved charges are given by
\begin{eqnarray}\label{dicke:conservedcharges}
& R_i(\xi)=S_i^0+g\sum_{k\neq 0,i}^m[\case{1}{2}X_{ik}(S_i^\dag S_k + S_k^\dag S_i)+Z_{ik} S_i^0 S_k^0]\nonumber\\
&\qquad+g[\case{1}{2}X_{i0}(S_i^\dag A_0(\xi)+A_0^\dag(\xi)S_i)+Z_{i0}A_0^0(\xi)S_i^0].
\end{eqnarray}
From here on, we will employ the trigonometric realization (\ref{rq:trigonometric}) of the Gaudin algebra (\ref{rg:gaudinalgebra}).  Without loss of generality, the $X_{0k}$ and $Z_{0k}$ matrix elements can be evaluated in the $\eta_0\rightarrow\infty$ limit, giving rise to the parametrization
\begin{equation}\label{dicke:x0kz0k}
X_{0k}=\lim_{\eta_0\rightarrow\infty}\case{\sqrt{1+\eta_0^2}\sqrt{1+\eta_k^2}}{\eta_0-\eta_k}=\sqrt{1+\eta_k^2},\qquad
Z_{0k}=\lim_{\eta_0\rightarrow\infty}\case{1+\eta_0\eta_k}{\eta_0-\eta_k}=\eta_k.
\end{equation}
Using this realization and the definition of the Gaudin algebra (\ref{rg:gaudinalgebra}), it is straightforward to show that 
\begin{equation}
X_{ik}=\frac{X_{i0}X_{0k}}{Z_{i0}+Z_{0k}}=\frac{\sqrt{1+\eta_i^2}\sqrt{1+\eta_k^2}}{\eta_i-\eta_k},\qquad
Z_{ik}=\sqrt{X_{ik}^2-1}=\frac{1+\eta_i\eta_k}{\eta_i-\eta_k},
\end{equation}
which is indeed a standard trigonometric parametrization (\ref{rq:trigonometric}) of the Gaudin algebra (\ref{rg:gaudinalgebra}).  The set of parameters $\{\eta_k\}$ can be chosen freely, so we renormalize them as $\eta_k=-\sqrt{\frac{2\xi}{\Omega_0G^2}}\varepsilon_k$, with $\{\varepsilon_k\}$ a set of free, $\xi$-independent, variables.  It is noteworthy that the $X_{0k}$ and $Z_{0k}$ matrix elements take the form
\begin{equation}
X_{0k}= 1+\case{\xi}{\Omega_0G^2}\varepsilon^2_k +\mathcal{O}(\xi^2),\qquad Z_{0k}=-\sqrt{\case{2\xi}{\Omega_0G^2}}\varepsilon_k,
\end{equation}
in the $\xi\rightarrow0$ contraction limit, which is related to the parametrization proposed by Dukelsky et.\ al.\ \cite{dukelsky:2004b} in their derivation of the Dicke model.  Substituting the expressions of $X_{0k}$ and $Z_{0k}$ (eqs.\ (\ref{dicke:x0kz0k})) in the conserved charge, and taking the proper $\xi\rightarrow0$ limit, one obtains the Hamiltonian of the Dicke model (up to a divergent constant)
\begin{equation}
H_{\textrm{Dicke}}=\hbar\omega R_0(\xi\rightarrow0)=\hbar\omega b^\dag b + \sum_{k=1}^m\varepsilon_kS_k^0 +G\sum_{k=1}^m(b^\dag S_k + S_k^\dag b),
\end{equation}
where the $i=0$ notation has been omitted for the bosonic operators.  A similar procedure leads to the other conserved charges
\begin{equation}
\hbar\omega R_i(\xi\rightarrow0)=(\hbar\omega-\varepsilon_i) S_i^0+\sum_{k\neq i}^m\frac{2G^2}{\varepsilon_{k}-\varepsilon_i}[\case{1}{2}(S_i^\dag S_k+S_k^\dag S_i)+S_i^0S_k^0]-G(S_i^\dag b+b^\dag S_i).
\end{equation}

To obtain an expression for the Bethe Ansatz state in the contraction limit, one needs a parametrization of the matrix elements $X_{0\alpha}$ and $X_{k\alpha}$.  The former can be chosen according to eqs.\ (\ref{dicke:x0kz0k}) with $\eta_{\alpha}=-\sqrt{\frac{2\xi}{\Omega_0G^2}}x_\alpha$ , from which the latter can be derived
\begin{equation}
X_{k\alpha}=\frac{X_{k0}X_{0\alpha}}{Z_{k0}+Z_{0\alpha}}=\case{\sqrt{1+\eta_k^2}\sqrt{1+\eta_\alpha^2}}{\eta_k-\eta_\alpha}.
\end{equation}
Substituting these parametrisations for $X_{0\alpha}$ and $X_{k\alpha}$ into the Bethe Ansatz state eq.\ (\ref{rg:betheansatz}), and taking the $\xi\rightarrow0$ limit, one obtains
\begin{equation}
|\psi\rangle=\left(\frac{\Omega_0}{2\xi}\right)^{\frac{N}{2}}\prod_{\alpha=1}^N\left(b^\dag-G\sum_{k=1}^m\frac{S_k^\dag}{\varepsilon_k-x_\alpha}\right)|\theta\rangle,
\end{equation}
where the factor $(\frac{\Omega_0}{2\xi})^{\frac{N}{2}}$ can be absorbed by the normalization of the state.  

Finally, the RG equations for the Dicke model can be obtained likewise.  Starting from the RG equations (\ref{rg:rgequations}) with the $i=0$ quasispin copy replaced by a pseudo deformed $su(2)$,
\begin{equation}
1+gZ_{0\alpha}s_0(\xi)+g\sum_{k=1}^m Z_{k\alpha}s_k-g\sum_{\beta\neq\alpha}Z_{\beta\alpha}=0,\qquad\forall \alpha=1\dots N.
\end{equation}
and substituting the proper expressions for $Z_{k\alpha}$ and $Z_{\beta\alpha}$
\begin{equation}
Z_{k\alpha}=\frac{1+\eta_k\eta_\alpha}{\eta_k-\eta_\alpha},\qquad Z_{\beta\alpha}=\frac{1+\eta_\beta\eta_\alpha}{\eta_\beta-\eta_\alpha},
\end{equation}
we obtain the RG equations for the Dicke Hamiltonian
\begin{equation}
(\hbar\omega-x_\alpha)-2G^2\sum_{k=1}^m\frac{s_k}{\varepsilon_k-x_\alpha}+2G^2\sum_{\beta\neq\alpha}^N\frac{1}{x_\beta-x_\alpha}=0,\quad\forall\alpha=1\dots N.
\end{equation}
\section{Conclusions}
In conclusion, we have derived the RG integrability of the Dicke model in the contraction limit of a single pseudo deformed quasispin algebra copy in the $\otimes_{i=0}su(2)_i$ RG system.  In addition to the Dicke Hamiltonian, the additional conserved charges for the integrability have been obtained, as well as the correct form for the Bethe Ansatz state and the RG equations which the rapidities in the Bethe Ansatz need to satisy.  It would be interesting to investigate how the Bethe Ansatz eigenstates of the trigonometric RG Hamiltonian are adiabatically connected to the Bethe Ansatz eigenstates of the Dicke model.   This will constitute the subject of future investigations.
\ack
PWC and SDB acknowledge financial support from FWO-Vlaanderen as pre-doctoral and post-doctoral fellows respectively.

\section*{References}


\providecommand{\newblock}{}

\end{document}